\documentstyle{article}\begin{document}
\title{Einstein gravity of a diffusing fluid  }
\author{ Z. Haba\\
Institute of Theoretical Physics, University of Wroclaw,\\ 50-204
Wroclaw, Plac Maxa Borna 9, Poland\\
email:zhab@ift.uni.wroc.pl}\maketitle
\begin{abstract}
We discuss Einstein gravity for a  fluid consisting of particles
interacting with an  environment of some other particles. The
environment is described by a time-dependent cosmological term
which is compensating the lack of the conservation law of the
energy-momentum of the dissipative fluid. The dissipation is
approximated by a relativistic  diffusion in the phase space. We
are interested in a homogeneous isotropic flat expanding Universe
described by a scale factor $a$. At an early stage the particles
are massless. We obtain explicit solutions of the diffusion
equation for a fluid of massless particles at finite temperature.
The solution is
 of the form of a modified
J\"uttner distribution with a time-dependent temperature. At later
time Universe  evolution is described as a diffusion at zero
temperature with no equilibration. We find solutions of the
diffusion equation at zero temperature which can be treated as a
continuation to a later time of the finite temperature solutions
describing an early stage of the Universe. The energy-momentum of
the diffusing particles is defined by their phase space
distribution. A conservation of the total energy-momentum
determines the cosmological term up to a constant. The resulting
energy-momentum inserted into Einstein equations gives a modified
Friedmann equation. Solutions of the Friedmann equation depend on
the initial value of the cosmological term. The large value of the
cosmological constant implies an exponential expansion. If the
initial conditions allow a power-like solution for a large time
then it must be of the form $a\simeq \tau$ (no deceleration,
$\tau$ is the cosmic time) . The final stage of the Universe
evolution is described by a non-relativistic diffusion of a cold
dust.

\end{abstract}
\section{Introduction}
We consider the gravitational Einstein equations in the form
\begin{equation} R^{\mu\nu}-\frac{1}{2}h^{\mu\nu}R=8\pi
G T^{\mu\nu},
\end{equation}
 where $G$ is the Newton constant, $h^{\mu\nu}$ is the metric and  the Einstein tensor on the
lhs is covariantly conserved. Hence,
\begin{equation}(T^{\mu\nu})_{;\mu}=0.\end{equation}
We could insert on  the rhs of eq.(1)  the energy-momentum
$T^{\mu\nu}$ of a collection of particles described by a probability
distribution $\Phi$ on the phase space
\cite{ellis}\cite{vega}\cite{pavlov}\cite{andrea}. If particle's
dynamics is determined by classical evolution equations, then the
conservation law (2) is a consequence of the Liouville equation
(where $\Gamma^{\mu}_{\nu\rho}$ are Christoffel symbols)
\begin{equation}
\begin{array}{l}
(p^{\mu}\partial^{x}_{\mu}-\Gamma^{k}_{\mu\nu}p^{\mu}p^{\nu}\partial_{k})\Phi=
0.
\end{array}\end{equation}
Eqs.(1),(3) and the formula for the energy-momentum tensor
\begin{equation}
T^{\mu\nu}=\sqrt{h}\int \frac{d{\bf
p}}{(2\pi)^{3}}\frac{1}{p_{0}}p^{\mu}p^{\nu}\Phi,
\end{equation} define the Einstein-Vlasov system
\cite{andrea}. In eq.(4)  $h$ is the determinant of the metric and
$p_{0}$ is determined from the mass-shell condition
$p_{\mu}p^{\mu}=m^{2}$ ($m$ is the particle's mass, we set $c=1$).
In eqs.(1)-(4) Greek indices run from $0$ to $3$, Latin indices
denoting spatial components have the range from $1$ to $3$, the
covariant derivative is over the space-time, derivatives over the
momenta $\frac{\partial}{\partial p^{k}}$ are denoted
$\partial_{k}$ and $\partial^{x}$ denotes a derivative over a
space-time coordinate $x$.

The Einstein-Vlasov system has been successfully applied
\cite{andrea} in a description of the large scale structure of the
Universe. The deterministic approach (1)-(4) must be modified if
we describe only a part of the total system. In such a case we do
not have the complete information, e.g., we do not detect some
forms of matter. The unknown part of the total system is excluded
from a deterministic description.  The system's motion is not
deterministic but rather random. Under the assumption of a short
memory (Markov approximation) the random motion can be described
as a diffusion. The non-relativistic diffusion approximation has
been widely applied in many body systems
\cite{pitaj}\cite{chandra}. We consider a relativistic diffusion
of the hot matter beginning from the early time of the Universe
expansion.  We apply the diffusion approximation to the
description of motion of the luminous matter in the Universe
assuming that this matter is formed from known elementary
particles (there is an alternative approach describing the dark
matter evolution as a diffusion \cite{berschinger}; for a
discussion of the dark matter phase space distribution see
\cite{dm}). We suggest that there are some other particles and
interactions which we do not know. Their impact is described as a
diffusive motion of matter.The gravitational attraction by distant
objects as a source of the diffusive motion of stars has been
described in \cite{chandra2} and confirmed by observations
\cite{wielen} \cite{henyey}. In this paper the diffusion
approximation is applied to all stages of the evolution.
 The energy-momentum tensor $T_{\mu\nu}$ on the rhs of eq.(1)
consists of two parts: the first part expressed in eq.(4) (which
is not conserved if $\Phi$ satisfies a diffusion equation) and the
second part corresponding to the unknown component of the total
system which is to be determined from the conservation law (2).

 We could consider a  Big Bang theory of a fluid consisting of
 particles of the standard grand unification model \cite{albrecht}\cite{dodelson}.
  The heavy ion collisions could supply   experiments
  verifying some assumptions of the Big Bang theory starting from
  microscopic kinetic theory \cite{dodelson}\cite{rafelski}\cite{svet}.  We are unable to treat the formation of hadrons from quarks,decay and creation of
 particles
 without
 a complete quantum theory.
 We  rely on a classical approximation of quantum theories.
 We approximate interactions by a diffusion. We have studied such an approximation in \cite{habajpa} where
  an interaction of relativistic  charged particles with the CMB radiation is treated
 as a diffusion model.
 In the same way we can show that
  the quark-gluon interaction leads to a diffusion of quarks \cite{habampa}.
  Even if the interactions are unknown the relativistic invariance
  determines to a large extent the form of the diffusion.
 The relativistic diffusion has been discovered in
 \cite{schay} and \cite{dudley}. A model with friction leading to
 the J\"uttner equilibrium distribution \cite{juttner} is discussed in \cite{habapre}
 \cite{calo}\cite{hang}\cite{chev}.
 An equilibration to a quantum distribution requires a non-linear diffusion
 \cite{habaphysa} which is to describe the bunching of Bose particles and repulsion of Fermi particles.
 We treat the linear diffusion as an approximation to the non-linear one.
  We think that the relativistic
 diffusion can be a useful approximation for dissipative processes
 in the description of the early Universe. The non-relativistic diffusion as an approximation
 for the collision term in the Boltzmann equation has been elaborated in many papers (see
 \cite{pitaj}\cite{chandra}\cite{henyey}; for a recent application in cosmology  see \cite{pavlov}
 and references cited there).  An equation for
the diffusion of photon energy in a medium of free
non-relativistic electrons is well-known as the Kompaneets
equation \cite{komp}\cite{wein}. A relativistic diffusion
approximation for the Boltzmann equation is discussed in
\cite{kremer}.

 In this paper we are going to elaborate the scheme suggesting that some unknown
 interactions  are approximated by
a relativistic diffusion. We assume that the particles are
massless (we discuss a diffusion of massless particles in
\cite{habaphysa} \cite{habampa2}\cite{habacqg}). In the grand
unified theories elementary particles acquire a mass through a
symmetry breaking at lower temperature. Hence, the zero mass
assumption at an early time (the radiation era) is justified.
 The CMB radiation
\cite{CMB} shows the quantum origin of the (Planck) equilibrium
distribution. We obtain a solution of the non-linear diffusion
equation in the form of the Planck distribution with a time
dependent temperature. In the ultrarelativistic limit and low
density the non-linear diffusion is approximated by the linear
diffusion. We solve the linear relativistic diffusion equation
exactly. We find initial states of the J\"uttner type
\cite{juttner} whose form is preserved by the time evolution. Such
(steady) states are analogs of the equilibrium states in an
expanding Universe. Moreover, they can be considered as asymptotic
phase space distributions in the sense that any solution of the
diffusion equation is approaching the J\"uttner distribution with
a time-dependent temperature.
  In a later time of the Universe evolution (lepton era) when other
kinds of matter appear with a different local temperature we do
not expect that the phase space distribution is approaching the
equilibrium distribution. We describe this later stage of  time
evolution of the energy-momentum  by a diffusive fluid of zero
temperature (no heat bath).In  order to perform exact calculations
we still apply an ultrarelativistic approximation setting the mass
of particles equal to zero .  The diffusion constant is defined by
the cross section of particle's interactions \cite{kremer}. It can
vary depending on the content of the system. We assume that this
variation is continuous so that we obtain a continuous transition
from the description in different stages of the Universe evolution
(we denote the diffusion constant at various stages of the
Universe expansion by the same symbol). The relativistic diffusion
at zero temperature with no equilibrium distribution has been
studied by Schay \cite{schay} and Dudley\cite{dudley}. It has been
applied for a description of a cosmological evolution of a
diffusing fluid  at a later stage of the expansion in \cite{calo1}
and \cite{calo2}. In this paper we discuss exact solutions of the
diffusion equation  in the massless limit applied to early stages
of the evolution. We obtain a continuation of the time-dependent
J\"uttner distribution at an early time to the solution of the
diffusion equation at zero temperature at later time. We insert
the energy-momentum corresponding to the solutions of the
diffusion equation into the Einstein equations (1)
\begin{equation}
R^{\mu\nu}-\frac{1}{2}h^{\mu\nu}R=T^{\mu\nu}=T_{D}^{\mu\nu}
+\tilde{T}^{\mu\nu},\end{equation} where $T_{D}$ is the
energy-momentum of a certain (dark) matter and $\tilde{T}$ is the
energy-momentum of the system of diffusing particles. From the lhs
it follows that
\begin{equation} (T^{\mu\nu}_{D})_{;\mu}=-
(\tilde{T}^{\mu\nu})_{;\mu}.\end{equation} Knowing the rhs of
eq.(6) we can determine the lhs up to a constant. We represent
$T_{D}$ by a time-dependent cosmological term $\Lambda$. The
cosmological term $\Lambda$ acquires a dynamical content in a way
similar to the trace free formulation of Einstein equations
\cite{weinberg}\cite{ellis2}. A dynamical relation of the
cosmological term to the matter  density seems to be unavoidable
for an explanation of the coincidence problem \cite{carrol}
\cite{strau}. In contradistinction to the Einstein-Vlasov system
the dissipative system (5) gives a relation which could answer the
question why the dark matter and "barionic" matter are of the same
order of magnitude.  We show that such a cosmological term is
decreasing in time starting from a constant at a fixed time
$\tau_{0}$ (which can be chosen as the radiation decoupling time).
After an addition of the cosmological term $T_{D}$ the total
energy-momentum $T$ can violate the strong energy condition as
well as  the dominant energy condition \cite{hawking}. In the case
of a homogeneous isotropic Universe we obtain an analog of the
Friedmann equation for the scale factor $a$. The solutions of the
modified Friedmann equations depend on the value of the
cosmological constant. They can grow exponentially.  We put the
particle's mass equal to zero in early stages of the evolution. At
the final stage of the expansion, when  matter has already a
macroscopic form, we consider a non-relativistic approximation (a
large mass limit) of the diffusion equation. We discuss the
resulting Friedmann equation for the scale factor $a$. We find an
exact solution of the non-relativistic diffusion equation starting
from the Maxwell equilibrium distribution. We insert the
corresponding energy-momentum tensor in Einstein equations. The
qualitative results for $a$ in the non-relativistic limit are
similar to those of the relativistic massless case.

  The plan of the paper is the following. In
sec.2 we review the relativistic   diffusion. In sec.3 we derive
solutions of the relativistic diffusion equation for massless
particles at finite temperature.  In sec.4 we consider the limit
of zero temperature of the solutions of sec.3. We discuss the
energy-momentum tensor on the rhs of Einstein equations in sec.5.
In sec.6 we obtain some consequences of the Friedmann equation
resulting from Einstein equations  with a diffusing fluid. A model
of non-relativistic diffusing particles of a large mass (a cold
"dust") is discussed in sec.7. The final section contains a
summary and an outlook.
\section{Relativistic  diffusion}
The diffusion  is usually treated as an approximation for the
collision term in the kinetic equation \cite{pitaj}. The diffusion
approximation for the relativistic kinetic equation \cite{cerci}
has not been developed in a sufficient detail (see however
\cite{kremer}). If we look for a relativistic generalization of
the Kramers diffusion defined on the phase space then it is
determined in the unique way by the requirement that the diffusing
particle moves on the mass-shell  ${\cal H}_{+}$  (see
\cite{schay}\cite{dudley}\cite{habapre}\cite{calo}\cite{habacqg}\cite{chevexpand})

\begin{equation}
 h_{\mu\nu}p^{\mu}p^{\nu}=m^{2}
 \end{equation}with the metric (flat expanding Universe)

\begin{equation}\begin{array}{l}
ds^{2}=h_{\mu\nu}dx^{\mu}dx^{\nu}=d\tau^{2}-\delta_{jk}a(\tau)^{2}dx^{j}dx^{k}
=a^{2}(\tau(t))(dt^{2}-d{\bf x}^{2}),\end{array}\end{equation} where
\begin{displaymath}
t=\int d\tau (a(\tau))^{-1}
\end{displaymath}
is the conformal time (we call $\tau$ the cosmic time; the notation
follows the textbook of Durrer \cite{durer}).

 The diffusion is generated by the Laplace-Beltrami operator on ${\cal H}_{+}$
\begin{equation}
\triangle_{H}=\frac{1}{\sqrt{g}}\partial_{j}g^{jk}\sqrt{g}\partial_{k}
\end{equation}
where
\begin{equation}
g^{jk}=m^{2}h^{jk}+p^{ j}p^{ k}
\end{equation}  $\partial_{j}=\frac{\partial}{\partial
p^{ j}}$ and $g=\det(g_{jk})$ is the determinant of  $g_{jk}$. The
operator $\triangle_{H}$ can be expressed in the form
\begin{equation}
\triangle_{H}=m^{2}a^{-2}\delta^{jk}\partial_{j}\partial_{k}+p^{
j}p^{ k}\partial_{j}\partial_{k} +3p^{ k}\partial_{k},
\end{equation}
where $k=1,2,3$.

The transport equation for the linear diffusion generated by
$\triangle_{H}$ reads
\begin{equation}
\begin{array}{l}
(p^{\mu}\partial^{x}_{\mu}-\Gamma^{k}_{\mu\nu}p^{\mu}p^{\nu}\partial_{k})\Omega=
\kappa^{2}\triangle_{H}\Omega,
\end{array}\end{equation} where $\kappa^{2}$ is the diffusion constant,
$\partial_{\mu}^{x}=\frac{\partial}{\partial x^{\mu}}$ and
$x=(t,{\bf x})$ (boldface letters denoting the three vectors).

The relativistic diffusion as an approximation of the relativistic
Boltzmann equation \cite{cerci} is discussed in ref.\cite{kremer}.
 It is shown in \cite{kremer} that  the diffusion constant $\kappa^{2}$ is
proportional to  the  cross section describing a scattering by
particles of the medium . The diffusion matrix $g^{jk}$ is expressed
by some integrals which have not been calculated explicitly in
\cite{kremer}.

Eqs.(9)-(10) and (12) hold true for a general metric $h_{\mu\nu} $
in the gauge $h_{0j}=0$ ( for a coordinate independent formulation
see \cite{franchi}). The specific realization of eq.(12) depends on
the use of either lower case
\cite{habacqg}\cite{calo1}\cite{calo2}\cite{chevexpand}, upper case
 \cite{franchi} or physical momenta ($\overline{p}=ap$)\cite{bernstein}. In this
 paper we use the upper case momenta ( the formula
 $p^{j}=h^{jk}p_{k}$ relates various formulations).

 We consider also a diffusion at finite
temperature $\beta^{-1}$\cite{habapre}\cite{calo}(in the conformal
time and in a frame moving with the heat bath)

\begin{equation}
\begin{array}{l}
(p^{\mu}\partial^{x}_{\mu}-\Gamma^{k}_{\mu\nu}p^{\mu}p^{\nu}\partial_{k})\Omega=
\kappa^{2}p^{0}\partial_{j}(g^{jk}\frac{1}{p^{
0}}\partial_{k}+\beta a^{2}p^{j})\Omega.
\end{array}\end{equation} Here, $p^{0}$
is determined from eq.(7). So, in the conformal time
\begin{equation} p^{0}=\sqrt{a^{-2}m^{2}+{\bf p}^{2}}
\end{equation}where
${\bf p}^{2}=\sum_{j}p^{j}p^{j}$.

In the metric (8)  the non-vanishing Christoffel symbols are
\begin{equation}
\Gamma^{0}_{jk}=a^{-1}\frac{da}{dt}\delta_{jk}
\end{equation}\begin{equation}
\Gamma^{j}_{0k}=a^{-1}\frac{da}{dt}\delta^{j}_{k}.
\end{equation}For rotation invariant
and space homogeneous $\Omega$ eq.(12) reads
\begin{equation}\begin{array}{l}
\sqrt{p^{2}+a^{-2}m^{2}}\Big(\partial_{t}\Omega-2{\cal
H}p\frac{\partial\Omega}{\partial
p}\Big)\cr=\kappa^{2}\Big(a^{-2}m^{2}\frac{\partial^{2}\Omega}{\partial
p^{2}}+2a^{-2}m^{2}p^{-1}\frac{\partial \Omega}{\partial
p}+p^{2}\frac{\partial^{2}\Omega}{\partial
p^{2}}+3p\frac{\partial\Omega}{\partial p} \Big)
\end{array}\end{equation}where  $p=\vert{\bf
p}\vert$ and \begin{equation} {\cal
H}=a^{-1}\frac{d}{dt}a=aH.\end{equation} Here, $H$ is the Hubble
expansion rate.

We have shown in \cite{habaphysa} that if the phase space
distribution has the Bose-Einstein or Fermi-Dirac equilibrium
limit which is a minimum of the relative entropy (related to the
free energy ) then the diffusion equation must be non-linear. The
proper generalization of eq.(13) (which leads to the equilibrium
in a static metric) reads

\begin{equation}
\begin{array}{l}
(p^{\mu}\partial^{x}_{\mu}-\Gamma^{k}_{\mu\nu}p^{\mu}p^{\nu}\partial_{k})\Omega=
\kappa^{2}p^{0}\partial_{j}\Big(g^{jk}\frac{1}{p^{
0}}\partial_{k}\Omega+\beta a^{2}p^{j}\Omega(1+\nu\Omega)\Big),
\end{array}\end{equation} where
$\nu=1$ for bosons and $\nu=-1$ for fermions.
 The classical (Boltzmann) statistics can be described by
 $\nu=0$. The relation to Kompaneets equation \cite{komp}\cite{wein}is discussed in
 \cite{habaphysa}.

In the ultrarelativistic limit  $m=0$ the dependence of the
diffusion generator (11) on the metric $h_{jk}$ disappears . In
such a case the quantum equilibrium distributions solve the
diffusion equation (19) in an expanding metric (8). In the case
$m=0$ and for rotation invariant functions eq.(19) is
\begin{equation}
\begin{array}{l}
\partial_{t}\Omega=2{\cal H}p\frac{\partial\Omega}{\partial
p}+\kappa^{2}p^{-2}\frac{\partial}{\partial
p}p^{3}(\frac{\partial\Omega}{\partial p}+ \beta
a^{2}\Omega(1+\nu\Omega)).
\end{array}\end{equation}
In the ultrarelativistic limit at low density the dependence on
the statistics should be irrelevant. We can neglect the quadratic
term in eq.(20) ($\nu=0$). Eq.(20) becomes linear\begin{equation}
\begin{array}{l}
\partial_{t}\Omega=\kappa^{2}p\frac{\partial^{2}\Omega}{\partial
p^{2}}+(3\kappa^{2}+2{\cal H}p+
\beta\kappa^{2}a^{2}p)\frac{\partial\Omega}{\partial p}
+3\beta\kappa^{2}a^{2}\Omega.
\end{array}\end{equation}In the initial stage of  the expansion  the frequent particles'
interactions allow to achieve an equilibrium. The quantum
equilibrium distribution will be a consequence of eq.(20). In the
ultrarelativistic regime the classical approximation should apply
leading to the J\"uttner equilibrium distribution \cite{juttner} .
 In fact,  the $\beta$-dependent term in eq.(21) can be obtained
from the principle of the detailed balance under the assumption that
at infinite time $\Omega$ should reach the J\"uttner equilibrium
distribution.

The assumption $m=0$ can be justified in grand unified theories at
high temperatures.  The approximation $p^{0}=\vert {\bf p}\vert$
at a later time can be treated as a zeroth order term in an
expansion of
  the
solutions of eqs.(12)-(13) in powers of $\frac{m}{ap}$. Eq.(17)
tends to the massless limit (21) when
$m^{2}a^{-2}p^{-2}\rightarrow 0$. In the  sequel we first restrict
ourselves to the massless (high energy) limit of both
eqs.(12)-(13). In sec.7 a non-relativistic limit (large mass
limit) of eq.(12) is applied for  a description of the final stage
of the Universe evolution as a cold dust.

Eq.(13) can be generalized to a non-isotropic Bianchi space-time
with the metric
\begin{displaymath}
ds^{2}=d\tau^{2}-a_{1}^{2}(dx^{1})^{2}-a_{2}^{2}(dx^{2})^{2}-a_{3}^{2}(dx^{3})^{2}\equiv
h_{jk}dx^{j}dx^{k}
\end{displaymath}where $a_{j} $ depend only on $\tau$.
The generalization (for massless particles) takes the form
\begin{equation}
\begin{array}{l}
(p^{\mu}\partial^{x}_{\mu}-\Gamma^{k}_{\mu\nu}p^{\mu}p^{\nu}\partial_{k})\Omega=
\kappa^{2}p^{0}\partial_{j}\Big(g^{jk}\frac{1}{p^{
0}}\partial_{k}+\beta
p^{j}\frac{1}{p^{0}}\sqrt{\sum_{l}p_{l}^{2}a_{l}^{4}}\Big)\Omega.
\end{array}\end{equation}where in the cosmic time
$p^{0}=\sqrt{h_{jk}p^{j}p^{k}}$.
\section{Solutions of linear and non-linear diffusion equations at finite temperature}
In a time-independent  metric any solution of the diffusion
equations (13) or (19) tends to an  equilibrium  which is a
time-independent solution of these equations. We are looking for
analogs of equilibrium solutions in the case of an expanding flat
Universe. It is easy to see that the time dependent Planck
distribution $\Omega_{E}^{PL}$ is the solution of eq.(20). We have
\begin{equation}
\Omega^{PL}_{E}=\Big(\exp(\beta a^{2}(p+\mu))-\nu\Big)^{-1}
\end{equation}
where $\mu$ is an arbitrary constant (the chemical potential). In
the ultrarelativistic limit ( a large $p$) the Planck distribution
(23) is the same as the J\"uttner distribution. However, the CMB
measurements \cite{CMB} show that the distribution (23) is valid
also for a small $p$. We are unable to derive other solutions
(starting from other initial phase space distributions) of
 the non-linear diffusion equation (20). Further on we discuss only
 the linear approximation (13)
 which can be considered as  the ultrareletivistic  limit of eq.(20).

We are looking for a general class of solutions of the linear
diffusion equation (13). We show that besides the time dependent
J\"uttner distribution there are more general solutions which
preserve their form under the time evolution. Hence, they could be
considered  as analogs of the equilibrium distribution. We show
that the  energy-momentum tensor (2) corresponding to such
solutions depends on the scale factor $a$ in a more complicated
way than it has been discussed in the literature so far ( see
\cite{adep}). The main conclusion of this section is that the
solution depends on an integral $A$ of $a^{2}$. In subsequent
sections we investigate the dependence of the energy-momentum on
$a$ and $A$. We show that the dependence on $A$ can be neglected
at a large time.

First, we look for solutions of eq.(21) in an exponential form
generalizing the J\"uttner distribution. Then, we show that an
initial distribution which is of the J\"uttner form multiplied by
a polynomial in $p$ preserves its shape under the time evolution.
Such momentum distributions could be considered as analogs of the
equilibrium distributions for a time-dependent metric. Assume the
solution of eq.(21) has the form
\begin{equation}
\Omega_{E}^{\beta}(t)=L(t)\exp(-\alpha(t)p).
\end{equation}
Then,
\begin{equation}
\partial_{t}L=-3\kappa^{2}\alpha L+3\beta \kappa^{2}a^{2}L
\end{equation}
and
\begin{equation}
-\partial_{t}\alpha=\kappa^{2}\alpha^{2}-2H\alpha-\beta\kappa^{2}a^{2}
\alpha.
\end{equation}
The solution of eq.(25) with the initial condition $L(t_{0})=1$ is
\begin{equation}
L=\exp(-3\kappa^{2}\int_{t_{0}}^{t}\alpha +3\beta \kappa^{2}A),
\end{equation}where
\begin{equation}
A(t)=\int_{t_{0}}^{t}a^{2}(\tau(r))dr.
\end{equation}
Let
\begin{equation}
w=\exp(\kappa^{2}\int \alpha)
\end{equation}
and $v=\partial_{t}w$. Then, from eq.(26)
\begin{equation}
v=ba^{2}(t)\exp(\beta \kappa^{2}A)
\end{equation}with a certain constant $b$.
Hence, from eq.(29)
\begin{equation}
\alpha=\beta a^{2}\exp(\kappa^{2}\beta A)\Big(R+\exp(\kappa^{2}\beta
A)\Big)^{-1}
\end{equation}
where from the initial condition we can determine $R$
\begin{equation}
\alpha(t_{0})=\beta a^{2}(t_{0})(R+1)^{-1}.
\end{equation}If the initial condition $\alpha(t_{0})=\beta a^{2} (t_{0})$
is chosen then $R=0$ and the
 solution (24) for any $t$ is
\begin{equation}
\Omega_{E}^{J}(t)=\exp(-\beta a^{2}p)\end{equation}(the J\"uttner
distribution \cite{juttner}). Moreover, for any initial condition if
$a(t)\rightarrow \infty$ when $t\rightarrow\infty$,  then for
$\Omega_{E}^{\beta}$ of eq.(24)
\begin{equation}\vert \Omega_{E}^{\beta} - \exp(- \beta
a^{2}(t)p)\vert\rightarrow 0
\end{equation} exponentially fast ( as  from eq.(31) $\alpha\rightarrow \beta a^{2}$
and from eq.(27) $L\rightarrow 1$).

Let
\begin{equation}
p=u^{2}.\end{equation} Then, eq.(21) takes the
form\begin{equation}\begin{array}{l}
\partial_{t}\Omega=\frac{\kappa^{2}}{4}\frac{\partial^{2}\Omega}{\partial
u^{2}}+\frac{5\kappa^{2}}{4}u^{-1}\frac{\partial\Omega}{\partial u}+
\frac{1}{2}(2{\cal H}+
\beta\kappa^{2}a^{2})u\frac{\partial\Omega}{\partial
u}+3\beta\kappa^{2}a^{2}\Omega.
\end{array}\end{equation}
In order to obtain another set of solutions  perturbing the
solution (33) we write
\begin{equation} \Omega=\Omega_{E}^{J}\Psi.
\end{equation}

We can express eq.(36) for $\Psi$ as an equation in 6 dimensional
momentum space ${\bf q}$ with ${\bf q}^{2}=u^{2}=p$ as follows
\begin{equation}\begin{array}{l}
\partial_{t}\Psi=\frac{\kappa^{2}}{4}\triangle_{6}\Psi+\frac{1}{2}\omega_{\beta}{\bf
q}\nabla\Psi,\end{array}\end{equation} where $\triangle_{6}$ is the
Laplacian in a 6 dimensional momentum space $R^{6}$ and
\begin{equation}
\omega_{\beta}=2{\cal H}- \beta\kappa^{2}a^{2}\end{equation}
Rotation invariant solutions of eq.(38) solve eqs.(36)-(37).
 In the space of
rotation invariant functions we can introduce a polynomial basis
\begin{equation}
\Psi(t,{\bf q})=1+\sum_{n=0}^{n=N} c_{n}(t)({\bf q}^{2})^{n}
\end{equation}
Inserting eq.(40) in eq.(38) we find linear equations for $c_{n}$
\begin{equation}
\partial_{t}c_{N}=N\omega_{\beta} c_{N},
\end{equation}\begin{equation}
\partial_{t}c_{n-1}=(n-1)\omega_{\beta} c_{n-1}+\kappa^{2} n(n+2)c_{n}
\end{equation} with $c_{0}(t_{0})=0$. These equations
can be solved systematically  starting from $N$.
 First
 \begin{equation}
c_{N}(t,t_{0})=c_{N}(t_{0})a(t)^{2N}a(t_{0})^{-2N}
\exp(-N\kappa^{2}\beta A(t)).
\end{equation}
Then, we can find the coefficients for lower $n$.
 In
detail, for $N=1$

\begin{equation}
c_{0}(t)=\frac{3}{\beta}a(t_{0})^{-2}\Big(1-\exp(-\kappa^{2}\beta
A(t))\Big)c_{1}(t_{0})
\end{equation}

and\begin{equation} c_{1}(t,t_{0})=c_{1}(t_{0})a^{2}(t)a(t_{0})^{-2}
\exp(-\kappa^{2}\beta A(t))\equiv \tilde{c}_{1}
a^{2}(t).\end{equation}

 For $N=2$ we have
\begin{equation}
c_{2}(t,t_{0})=c_{2}(t_{0})\frac{a(t)^{4}}{a(t_{0})^{4}}\exp(-2\kappa^{2}\beta
A)\equiv\tilde{c}_{2}a^{4},\end{equation}
\begin{equation}\begin{array}{l}
c_{1}(t,t_{0})=\frac{a^{2}(t)}{a^{2}(t_{0})}\exp(-\beta\kappa^{2}A)c_{1}(t_{0})\cr
+\frac{8}{\beta}\frac{a^{2}(t)}{a^{4}(t_{0})}
\exp(-\beta\kappa^{2}A)(1-\exp(-\beta\kappa^{2}A))c_{2}(t_{0})\equiv
\tilde{c}_{1}a^{2},\end{array}\end{equation}
\begin{displaymath}\begin{array}{l}
c_{0}(t,t_{0})=\frac{12}{\beta^{2}a^{4}(t_{0})}
(1-\exp(-\beta\kappa^{2}A))^{2}c_{2}(t_{0}).\end{array}\end{displaymath}
From eqs.(41)-(42) we can derive the behaviour of $ c_{n-1}$ as
$\simeq a^{2(n-1)}$. In fact, we have
\begin{displaymath}
\begin{array}{l}
c_{n-1}(t)=\exp(\int_{t_{0}}^{t}(n-1) \omega_{\beta} )c_{n-1}(t_{0})
+\kappa^{2}n(n+2)\int_{t_{0}}^{t}\exp(\int_{r}^{t}(n-1)\omega_{\beta})
c_{n}(r)dr.
\end{array}\end{displaymath}
Now, if $c_{n}\simeq a^{2n}$ then
$\exp((n-1)\int_{t_{0}}^{t}\omega_{\beta})\simeq
a(t)^{2(n-1)}a(t_{0})^{-2(n-1)}$ and
$\exp((n-1)\int_{r}^{t}\omega_{\beta})c_{n}(r)\simeq
a(t)^{2(n-1)}a^{2}(r)$. It follows that $c_{n-1}(t)\simeq
a^{2(n-1)}(t) f( A(t))$ with a certain bounded function $f$.
Hence, from the $c_{n}\simeq a^{2n}$ behaviour  we obtain the
behaviour $c_{n-1}\simeq a^{2(n-1)}$ and inductively for any lower
$n$. It can also be seen from eqs.(41)-(42) that  we have
$c_{n}\rightarrow 0$ (for $n>0$) because these coefficients
($\simeq f(A)$) decay exponentially with a growing $A$ and a
multiplication by powers of $a$ does not change the limit
$t\rightarrow \infty$.

The diffusion equation (22) for the Bianchi space-time has a
$\kappa$-independent solution $\Omega_{E}^{B}$ which is an analog
of the J\"uttner solution (33)
\begin{displaymath}
\Omega_{E}^{B}=\exp\Big(-\beta\sqrt{a_{1}^{4}p_{1}^{2}+a_{2}^{4}p_{2}^{2}+a_{3}^{4}p_{3}^{2}}\Big).
\end{displaymath}
However, in the anisotropic case we are unable to find analogs of
the $\kappa$-dependent solutions which we have discussed in this
section and the ones to be considered in the next section.
Nevertheless, the distribution $\Omega_{E}^{B}$ is interesting as
an example leading to  an anisotropic energy-momentum.

\section{A transition from finite temperature to zero temperature}
There are complex quantum processes involved in the early
Universe. We have at our disposal classical Einstein equations. We
assume that the momentum distribution of particles in the Universe
soon after the Big Bang can be described by the non-linear
diffusion equation for massless particles. We suggest that (like
in the linear case (34)) the solution of the non-linear equation
starting from an arbitrary initial condition is approaching for a
later time the time-dependent Planck distribution. In an early
stage of the Universe evolution  eq.(19) describes the diffusive
behaviour of the phase space distribution leading to the
 Planck  distribution (23). In the ultrarelativistic approximation we restrict ourselves to
the linear equation (19) with the J\"uttner equilibrium
distribution. After some time $t_{0}$ (beginning of the lepton
era) new forms of matter are created
  weakly interacting with radiation.
 We do not expect that J\"uttner phase space  distribution remains valid for
  all kinds of matter at a later time. We shall
describe the subsequent time evolution beginning at $t=t_{0}$ at
the expansion scale $a(t_{0})$ and at the temperature
$\beta^{-1}=\kappa^{2}\theta(a^{2}(t_{0}))^{-1}$ by the diffusion
equation (12) corresponding to a diffusion at zero temperature
with  no equilibration. The solutions of eq.(21) on the time
interval $[0,t_{0})$ can be transferred continuously to solutions
of eq.(12) on the interval $[t_{0},\infty)$ .  The diffusion
constant $\kappa$ in eq.(21) is determined by scattering processes
of the hot matter in the early Universe. It is not the same as the
diffusion constant in eq.(12) relevant for a later time when
scattering processes of different particles formed at later time
of the evolution are involved.

In this section we follow the procedure of the previous section
finding first a solution of the diffusion equation (12) of an
exponential (J\"uttner) type. Subsequently, we obtain analogs of
the solutions (37) obtained by a polynomial perturbation of the
exponential solution. The parameter $\beta$ in eq.(21) is to
disappear at $t=t_{0}$ when the phase space distribution begins to
evolve according to eq.(12). If $\beta \rightarrow 0$ in eq.(31)
then (with $1+R=\theta\beta\kappa^{2}$ ) we obtain
\begin{equation}
\alpha=\kappa^{-2} a^{2}(\theta+ A)^{-1}.
\end{equation}In this way we find
a particular solution of eq.(21)($\beta=0$) starting from the
J\"uttner distribution  $\Omega_{0}$ at
$t=t_{0}$\begin{displaymath}
\Omega_{0}=\exp(-(\kappa^{2}\theta)^{-1}a^{2}(t_{0})p)
\end{displaymath}  (the parameter
$\kappa^{2}\theta(a^{2}(t_{0}))^{-1}$ has the meaning of
temperature) then continuing at  $t\geq t_{0}$ as
\begin{equation}
\Omega_{E}(t)=\theta^{3}(\theta+A)^{-3}\exp(-\kappa^{-2}\frac{a^{2}}{\theta+A}p).
\end{equation} $\Omega_{E}(t)$ solves eq.(12) (eq.(21) with $\beta=0$) with the initial condition $\Omega_{0}$.

If the diffusion (12) starts from the Planck distribution (23)
($\mu=0$) then it continues as
\begin{equation}
\begin{array}{l}
\Omega_{E}^{P}(t)=\sum_{k=1}^{\infty}\nu^{k-1}(1+\kappa^{2}k\beta
A)^{-3}\exp(-\frac{a^{2}\beta k}{1+\kappa^{2}k\beta A}p)
\end{array}\end{equation}At large momenta the terms with $k>1$
in eq.(50)  are negligible. We treat the J\"uttner-type solution
(49) at zero temperature as the high-energy approximation to the
Planck-type solution (50).

In the J\"uttner case we can  insert again eq.(37) into eq.(36)
with $\beta=0$ and derive eqs.(41)-(42) with

\begin{equation}
\omega=-\frac{d}{dt}\ln (\frac{\theta+A}{a^{2}})\end{equation} We
have solved eqs.(41)-(42) explicitly for small $N$. We have

 \begin{equation}
c_{N}(t,t_{0})=c_{N}(t_{0})a(t)^{2N}a(t_{0})^{-2N}
(\theta+A)^{-N}\theta^{N}.\end{equation} Then, for $N=1$
\begin{equation}
c_{1}(t)=\frac{a^{2}(t)}{a^{2}(t_{0})}(\theta+A)^{-1}\theta
c_{1}(t_{0})\equiv \tilde{c}_{1}a^{2}(t)(\theta+A)^{-1},
\end{equation}
\begin{equation}
c_{0}(t)=3c_{1}(t_{0})\kappa^{2}\theta\ln(1+\frac{A}{\theta}).
 \end{equation}
 For $N=2$ we have
\begin{equation}
c_{2}(t)=\frac{a(t)^{4}}{a^{4}(t_{0})}(\theta+A)^{-2}\theta^{2}c_{2}(t_{0})
\equiv\tilde{c}_{2}a^{4}(t)(\theta+A)^{-2},\end{equation}

\begin{equation}\begin{array}{l}
c_{1}(t)=\frac{a^{2}(t)}{a^{2}(t_{0})}(\theta+A)^{-1}\theta
c_{1}(t_{0}) +
8c_{2}(t_{0})\kappa^{2}a^{2}(t)a(t_{0})^{-4}\theta^{2}(\theta+A)^{-1}\ln(1+\frac{A}{\theta})\cr
\equiv \tilde{c}_{1}a^{2}(t)(\theta+A)^{-1},
\end{array}
\end{equation}
\begin{equation}\begin{array}{l}
c_{0}(t)=3c_{1}(t_{0})\kappa^{2}a(t_{0})^{-2}\theta\ln(1+\frac{A}{\theta})
+24\kappa^{4}\theta^{2}c_{2}(t_{0})a(t_{0})^{-4}(\ln(1+\frac{A}{\theta}))^{2}.
\end{array}\end{equation}
The solutions (52)-(57) will be discussed in subsequent sections.

\section{Einstein equations for diffusing particles at high temperature}
In the remaining sections we discuss the Einstein equations (1).
In principle, we should  discuss all fields and particles entering
the energy-momentum and consider a quantum version of Einstein
equations. However, to be realistic we accept the point of view
that we are far from a knowledge of all forms of matter and energy
and their interaction. Having a partial knowledge we approximate
the impact of the unknown interactions by a diffusion whereas the
energy-momentum of unknown particles and interactions is described
by $T_{D}^{\mu\nu}$ in eq.(5). The conservation of the total
energy-momentum determines $T_{D}$ in terms of the energy-momentum
$\tilde{T}$ of diffusing particles according to eq.(6). An
indication of a necessity of such an approach is the coincidence
problem \cite{strau}(the dark matter and luminous matter are of
the same order). The main aim of this section is a derivation of
$T_{D}$ from eq.(6). For a homogeneous and isotropic Universe this
will allow us to write down the Einstein equations (5) for the
scale factor $a$ (the Friedmann equations) explicitly. We shall
derive the dependence of the energy-momentum on the scale factor
$a$ using the solutions of the diffusion equation obtained in
secs.3 and 4. We discuss the behaviour of
 solutions of the Friedmann equations from the point of view of
 the energy conditions satisfied by $\tilde{T}+T_{D}$.

We calculate the energy-momentum using the phase space
distribution resulting from the solution of the diffusion
equation. In the early Universe this should be the non-linear
diffusion leading to the experimentally confirmed \cite{CMB} CMB
distribution (23). In eq.(4) we have already inserted the division
of the quantum phase space into elementary units of the volume
$(2\pi\hbar)^{3}$ ( we set $\hbar=1$). The solutions (23) and (33)
are expressed in the comoving frame (moving with the fluid). We
restrict ourselves in this section to the linear approximation of
the diffusion equation (20) at finite temperature ( without the
heat bath we have always the linear equation (12)).
 If in a new frame the fluid is moving with a velocity $w$ ($w^{\mu}w_{\mu}=1$ ) then
in this frame the time is $s=w_{\mu}x^{\mu}$ and the energy
$p^{\mu}w_{\mu}$. The scale factor $a$ depends on $s$.
$\Omega_{E}^{J}$ (33) is expressed in a covariant form as
\begin{equation} \Omega^{J}_{E,n}=nM^{-1}\exp(-\beta
a(s)w_{\mu}p^{\mu}),
\end{equation} where $n$ and $M$ are normalization factors.
The normalization is introduced \cite{norma} according to the rule
that the normalized state $\Omega_{n}$  satisfies the condition
\begin{equation}\sqrt{h}\int \frac{d{\bf
p}}{(2\pi)^{3}}\Omega_{n}=n.
\end{equation}
Here, $n$ is the density defined by the current
\cite{dodelson}\cite{durer}
\begin{equation}
N^{\mu}\equiv nu^{\mu}=\sqrt{h}\int \frac{d{\bf
p}}{(2\pi)^{3}}\frac{1}{p_{0}}p^{\mu}\Omega_{n}.
\end{equation}which determines the density $n$ and the velocity $u$ of the fluid ( $u^{\mu}u_{\mu}=1$).
If for an unnormalized state $\Omega$ we define
\begin{equation}M=\sqrt{h}\int \frac{d{\bf
p}}{(2\pi)^{3}}\Omega,
\end{equation}
then the normalization of an arbitrary distribution $\Omega $ is

 \begin{equation}
 \Omega_{n}=nM^{-1}\Omega.
 \end{equation}
 We define the  energy-momentum tensor as
\cite{dodelson} \cite{durer}
\begin{equation}
\tilde{T}^{\mu\nu}=\sqrt{h}\int \frac{d{\bf
p}}{(2\pi)^{3}}\frac{1}{p_{0}}p^{\mu}p^{\nu}\Omega_{n}.
\end{equation}
 We are going to study the Universe
evolution described by the states (37) either for a large or for a
small time. It is usually discussed in the cosmic time $\tau$.
Note that in conformal time $\sqrt{h}=a^{4}$, $p_{0}=\vert{\bf
p}\vert$, whereas in cosmic time $\sqrt{h}=a^{3}$ and
$p_{0}=a\vert{\bf p}\vert$. From now on we express all equations
in the cosmic time. If there is no diffusion then $\Omega$
satisfies the Liouville equation (3). The energy-momentum tensor
is conserved. This is the case for $\Omega_{E}$ of eqs.(23) and
(33). In this state both sides of eq.(13) are equal to  zero.
Hence, eq.(3) is satisfied. The formula (63)
 gives  $\tilde{T}^{00}\simeq a^{-4}$ in the states (23) and (33). In the
  J\"uttner state (23)
  we obtain from eq.(1) the  Friedmann equation (flat space, cosmic time)
\begin{equation}
(a^{-1}\frac{da}{d\tau} )^{2}=\frac{8\pi
G}{3}\frac{1}{(2\pi)^{3}}24\pi (\beta a)^{-4}.
\end{equation}
In the Planck  state (33) there will be an extra factor
$\frac{\pi^{4}}{90}\simeq 1$ on the rhs of eq.(64) corresponding
to the correction for small momenta contribution to the integral
(63). The resulting Friedmann equation coincides with the standard
one (see \cite{durer}, sec.1.3).

The energy-momentum tensor (63) is not conserved if the solution
of the diffusion equation depends on $\kappa$ as the general
solutions of secs.3 and 4 do. In general, for massless particles
we have \cite{habampa}\cite{calo1}
\begin{equation}
(\tilde{T}^{\mu\nu})_{;\mu}=3\kappa^{2}N^{\nu}-\beta\kappa^{2}\tilde{T}^{0\nu}.
\end{equation}
Using the Christoffel symbols (15)-(16) the covariant divergence of
an arbitrary tensor $T$ reads \begin{displaymath} (T^{\mu
0})_{;\mu}=\partial_{\tau}T^{00}+3a\frac{da}{d\tau}T^{00}+a\frac{da}{d\tau}\delta_{jk}T^{jk}.
\end{displaymath}
In a homogeneous Universe
\begin{equation}
\tilde{T}^{\mu\nu}=\tilde{E}u^{\mu}u^{\nu}-\tilde{\pi}_{E}(h^{\mu\nu}-u^{\mu}u^{\nu}),
\end{equation}
where $\tilde{E}$ is the energy,  $\tilde{\pi}_{E}$ the pressure
and the four-velocity $u^{\mu}$ satisfies the condition
\begin{equation}
h_{\mu\nu}u^{\mu}u^{\nu}=1.
\end{equation}
 For massless particles
$\tilde{T}^{\mu}_{\mu}=0$. Hence,
\begin{equation}
\tilde{\pi}_{E}=\frac{1}{3}\tilde{E}.
\end{equation}
In the frame $u=(1,{\bf 0})$ we have
\begin{equation}\begin{array}{l} (\tilde{T}^{\mu
0})_{;\mu}=\partial_{\tau}\tilde{E}+3 a^{-1}\partial_{\tau}a
(\tilde{E}+\tilde{\pi}_{E}).\end{array}\end{equation} We shall
represent the unknown energy $T_{D}$ in eq.(5) by a cosmological
term $\Lambda$. Then
\begin{equation}
T^{\mu\nu}=\tilde{T}^{\mu\nu}+h^{\mu\nu}\frac{\Lambda}{8\pi G},
\end{equation} where $\tilde{T}$ is the energy-momentum tensor (63). The energy conservation (2) is expressed as
\begin{equation}\begin{array}{l} -\partial_{\tau}\frac{\Lambda}{8\pi G}=\partial_{\tau}\tilde{E}+3
a^{-1}\partial_{\tau}a
(\tilde{E}+\tilde{\pi}_{E})\end{array}\end{equation} In the
massless case (68) eq.(71) gives\begin{equation}
\frac{\Lambda(\tau)}{8\pi G}=\frac{\Lambda(\tau_{0})}{8\pi
G}-a^{-4}\int_{\tau_{0}}^{\tau}\partial_{r}(a^{4}\tilde{T}^{00})dr\end{equation}
 $\Lambda $ is determined from
eq.(71) up to a constant. We could also integrate eq.(65) in order
to express $\Lambda$ by $\tilde{T}$ and $N^{\mu}$. From eq.(65) it
can be seen that the divergence of $\tilde{T}^{\mu\nu}$ is
proportional to the diffusion constant. Applying eq.(72) we obtain
the Einstein equations (for a flat Universe) in the form
\begin{equation}\begin{array}{l} \frac{3}{8\pi G}H^{2}\equiv\frac{3}{8\pi G}(a^{-1}\frac{da}{d\tau}
)^{2}=\tilde{T}^{00}(\tau)-\int_{\tau_{0}}^{\tau}dra^{-4}\partial_{r}(a^{4}\tilde{T}^{00})
+\frac{\Lambda}{8\pi G}(\tau_{0})\cr
=\tilde{T}^{00}(\tau_{0})-4\int_{\tau_{0}}^{\tau}drH(r)\tilde{T}^{00}(r)+\frac{\Lambda}{8\pi
G}(\tau_{0})\end{array}\end{equation} It follows from eq.(73) that
the modification of the Friedmann equation resulting from the
diffusion is determined by the deviation of $\tilde{T}^{00}$ from
the standard $a^{-4}$ (as in eq.(64)).

 Eq.(73) can
also be expressed in the form
\begin{equation}\begin{array}{l}  \frac{3}{8\pi G}(H^{2}(\tau)-  H^{2}(\tau_{0}))=
-4\int_{\tau_{0}}^{\tau}dr H(r)\tilde{T}^{00}(r),
\end{array}\end{equation}
where we have used the relation
\begin{displaymath}\begin{array}{l}  \frac{3}{8\pi G} H^{2}(\tau_{0})
=\tilde{T}^{00}(\tau_{0})+\frac{\Lambda}{8\pi G}(\tau_{0})
=\tilde{E}(\tau_{0})+\frac{\Lambda}{8\pi G}
(\tau_{0})\end{array}\end{displaymath}resulting from the notation
of eqs.(66) and (70)
\begin{equation}
T^{00}=\tilde{E}+\frac{\Lambda}{8\pi G} .\end{equation} If we know
$H$ at a certain moment $\tau_{0}$ then eq.(74) determines
$a(\tau)$ for a later time. Note that if $\frac{da}{d\tau}\geq 0$
for $\tau\geq \tau_{0}$ then from eq.(73) it follows that with
\begin{equation}
 \frac{3}{8\pi G}\lambda\equiv \tilde{E}(\tau_{0})+\frac{\Lambda}{8\pi G}(\tau_{0})-4\int_{\tau_{0}}^{\infty}drH(r)\tilde{T}^{00}(r)
\end{equation}we have the bounds
\begin{displaymath}
\begin{array}{l}
\lambda \leq (a^{-1}\frac{da}{d\tau} )^{2}\leq
 \frac{8\pi G}{3}\tilde{E}(\tau_{0})+\frac{\Lambda}{3}(\tau_{0})
\end{array}\end{displaymath}
If $\lambda $  of eq.(76) is non-negative then we obtain the
inequality (Gronwall inequality)
\begin{equation}
\exp(\tau  H(\tau_{0}))\geq a(\tau)\geq
\exp(\sqrt{\lambda}\tau)-\exp(\sqrt{\lambda}\tau_{0}).
\end{equation}
There can be an exponential behaviour in a certain time interval
when the cosmological constant dominates and a power-like
behaviour if the cosmological constant is cancelled by other terms
on the rhs of eq.(73).

It is useful to study the behaviour of solutions of Einstein
equations from the point of view of the energy conditions
\cite{hawking} \cite{wec}. We can represent $T^{\mu\nu}$ again in
the hydrodynamic form (66) (with $E$ and $\Pi_{E}$ without
tilde).We have already assumed that we have as a solution a flat
Universe. This is possible only if
$E=T^{00}=\tilde{T}^{00}+\frac{\Lambda}{8\pi G}\geq 0$ .  In the
hydrodynamic representation $\tilde{\Pi}=\frac{1}{3}\tilde{E}$
because $p^{2}=0$ (eq.(68)). We can conclude from eqs.(66) and
(70) that
\begin{equation} E+3\Pi_{E}=2(\tilde{E}-\frac{\Lambda}{8\pi G})
\end{equation}
We have
$E+\Pi_{E}=\tilde{E}+\tilde{\Pi}_{E}$,$\Pi_{E}=\tilde{\Pi}_{E}-\frac{\Lambda}{8\pi
 G}$ and $E=\tilde{E}+\frac{\Lambda}{8\pi G}$. From eq.(78) we can
see that the strong energy condition (requiring that (78) be
positive \cite{hawking})can be violated if the cosmological term
is present. Using the definition of $\lambda$ (eq.(76)) we can
obtain (under the assumption $\partial_{\tau}a\geq 0$ for
$\tau\geq \tau_{0}$)
\begin{equation}
\tilde{E}(\tau)-\frac{\Lambda(\tau)}{8\pi G}\leq 2\tilde{E}(\tau)
-\frac{3\lambda}{8\pi G}
\end{equation}
In cosmological models the energy density $\tilde{E}(\tau)$ is
decreasing to zero when the Universe expands. Hence, if  $\lambda$
is positive  then according to eqs.(78)-(79) the strong  energy
condition will be violated for a sufficiently large time. From
Einstein equations $\partial_{\tau}^{2}a\simeq -a(E+3\Pi_{E})$.
Hence, (78) will be negative if the expansion  accelerates. This
explains the role played by the positivity of $\lambda$ in
eqs.(77) and (78).

It is clear that the weak energy condition $E\geq 0$ and
$E+\Pi_{E}\geq 0$ is satisfied. The dominant energy condition is
equivalent to $E=\tilde{E}+\frac{\Lambda}{8\pi G}\geq \vert
\Pi_{E}\vert =\vert \frac{1}{3}\tilde{E}-\frac{\Lambda}{8\pi G}\vert
$. Using eq.(72) this inequality can be rewritten as either
$\Lambda\geq -\frac{8\pi G}{3}\tilde{E}(\tau)$ or equivalently as
\begin{displaymath}
\tilde{E}(\tau_{0})+\frac{\Lambda(\tau_{0})}{8\pi G}\geq
\frac{2}{3}\tilde{E}(\tau) +4\int_{\tau_{0}}^{\tau}dr
H(r)\tilde{T}^{00}(r).
\end{displaymath}
In cosmological models $\tilde{E}(\tau)$ is decreasing to zero for
a large time. Hence, we can  obtain a sufficient condition for the
dominant  energy condition to be satisfied for a sufficiently
large $\tau$
\begin{equation}
\tilde{E}(\tau_{0})+\frac{\Lambda(\tau_{0})}{8\pi G}-
4\int_{\tau_{0}}^{\infty}dr H(r)\tilde{T}^{00}(r)=\frac{3}{8\pi
G}\lambda>0.
\end{equation}If $\lambda >0$ then the strong energy condition
is violated whereas the dominant energy condition is satisfied. It
can be seen that both conditions depend on the initial value for
the diffusion phase space distribution and the initial value of
the cosmological term.

Einstein equations  with the energy-momentum calculated in the
states at finite temperature are applied only at small time.  If
$a(\tau)$ tends to zero at small $\tau$ then we can derive the
behaviour  of $a(\tau) $ from the behaviour of the energy-momentum
for a small $a$.
 We calculate the
energy-momentum in the states of sec.3. For the state (31) we
obtain (here $\sqrt{h}=h^{3}$)
\begin{equation}
\tilde{T}^{00}=L(t)\sqrt{h}\int \frac{d{\bf
p}}{(2\pi)^{3}}ap\exp(-\alpha(t)p)=L(t)\frac{1}{(2\pi)^{3}}24\pi
(\beta a)^{-4}\Big(1+R\exp(-\kappa^{2}\beta A)\Big)^{4}
\end{equation}
For the state (44)-(45) we have\begin{equation}
\tilde{T}^{00}=\frac{1}{(2\pi)^{3}}\Big(24\pi (\beta
a)^{-4}(1+c_{0}(1))+96\pi(\beta a)^{-5}\tilde{c}_{1}(1)\Big)
\end{equation}and for (46)-(47)
\begin{equation}\begin{array}{l}
\tilde{T}^{00}=\frac{1}{(2\pi)^{3}}\Big(24\pi (\beta
a)^{-4}(1+c_{0}(2))+96\pi(\beta a)^{-5}\tilde{c}_{1}(2)\cr+ 480
\tilde{c}_{2}(2)(\beta a)^{-6}\Big)
\end{array}\end{equation}
We can calculate the remaining components of $\tilde{T}^{\mu\nu}$
 in the comoving frame $u=(1,0)$. We have according to eq.(66)
$\tilde{E}=\tilde{T}^{00}$, $\tilde{T}^{0j}=0$ and owing to the
rotational symmetry
$\tilde{T}^{jk}=\frac{1}{3}\delta^{jk}\tilde{T}^{00}$.  Then, from
eq.(68) $\tilde{\Pi}_{E}=\frac{1}{3}\tilde{T}^{00}$  in the
hydrodynamic representation (66) .

 The coefficients in eqs.(82)-(83) (where $c(2)$
means that this is $c$ at $N=2$)  depend on $A$ in such a way that
if $A$ tends to infinity at large time then the coefficients
$\tilde{c}_{1}$ and $\tilde{c}_{2}$ tend to zero. Hence, we do not
expect a modification of the large time behaviour at non-zero
temperature in the states (37). The argument at the end of sec.3
shows that the results (81)-(83) can be extended to general $N$
\begin{equation}
\tilde{T}^{00}= (\beta a)^{-4}\sum_{n=0}^{n=N}\tilde{c}_{n}(\beta
a)^{-n}
\end{equation}with $\tilde{c}_{n}$ (for $n>0$) decaying as $\exp(-\beta\kappa^{2}A)$.
For general $N$ we obtain again the conclusion that the polynomial
modification (37) of the J\"uttner distribution does not change
the large time behaviour of the energy-momentum in the Einstein
equations
 (note however that according to sec.4 at sufficiently large time the temperature
states are replaced by (49), their energy-momentum is discussed in
the next section). The dissipative modification (73) of the
Friedmann equations comes from a departure of
$a^{4}\tilde{T}^{00}$ from a constant (the constant is obtained
for a radiation without dissipation). From eq.(72) it follows that
$\Lambda $ is decreasing if $a^{4}\tilde{T}^{00}$ is increasing
(as will happen in the model (85) in the next section). The time
dependent terms $\tilde{c}_{n}$ disappear at large time but they
can be relevant for a rapid increase of $a(\tau)$ at an
intermediate time. The terms $\tilde{c}_{n}a^{-n}$ in eq.(84)
modify the short time behaviour of solutions of the Friedmann
equations as will be shown at the end of the next section.

\section{Universe expansion at zero temperature}
In our approach the energy-momentum tensor is defined by the
solution of the diffusion equation. We use different diffusion
equations at various stages of the expansion of the Universe. In
addition, the solution of the diffusion equation depends on the
initial condition.  The solutions of these equations determine
different energy-momenta. In a static metric a solution of the
diffusion equation tends to the unique equilibrium. The
energy-momentum tends to the one calculated in the equilibrium. In
a time-dependent metric strictly speaking an equilibrium solution
does not exist. However, we can distinguish some initial phase
space distributions which preserve their form during the time
evolution. We could also require that at fixed time such states
coincide with the static equilibrium states resulting from
variational principles of statistical mechanics (as is the case
with the Planck distribution $\Omega_{E}^{PL}$
 (23) and the J\"uttner distribution
$\Omega_{E}^{J}$ (33)). The distributions $\Omega_{E}^{PL}$ and
$\Omega_{E}^{J}$ lead to the standard dependence of the
energy-momentum $\tilde{T}\simeq a^{-4}$ on $a$ (in deterministic
models it also follows from the continuity equation and the
equation of state $E=w\Pi_{E}$, in the model (70) $w$ is not a
constant; for a discussion of $a$-dependence of $\tilde{T}$ see
\cite{adep}) . We have seen in sec.5 that the modified J\"uttner
distribution leads to the modified energy momentum depending on
$a$ as well as on $A$. However, the dependence on $A$ was
irrelevant at large time. The solutions of the diffusion equation
at finite temperature apply to an early stage of the evolution of
the Universe. At later time $\tau\geq\tau_{0}$ (after the
recombination time) the diffusion tending to an equilibrium cannot
correctly describe the phase space distribution of particles which
decouple from photons. In sec.4 we have described a continuation
of the evolution starting from the J\"uttner (33)(or Planck (23))
state and continuing according to the dissipative dynamics (12) of
the diffusion at zero temperature.
 In this section we calculate the energy-momentum $\tilde{T}$ of
  the  solution of the diffusion equation (12). Subsequently, from the energy-momentum conservation
  we determine the cosmological term $T_{D}$.
  We derive the Friedmann equation and find
  a  particular solution of this equation which is linear in time.
At the end of this section we return to the finite temperature
solutions of sec.5 in order to determine their behaviour at $a=0$.

  In the zero temperature state
(49) we obtain
\begin{equation}\begin{array}{l}
\tilde{T}^{00}=\sqrt{h}\theta^{3}(\theta+A)^{-3}\int \frac{d{\bf
p}}{2\pi)^{3}}ap\exp(-\kappa^{-2}\frac{a^{2}}{\theta+A}p)=
\frac{1}{(2\pi)^{3}}24\pi\kappa^{8}\theta^{3}(\theta+A)a^{-4}.\end{array}
\end{equation}
For the state (53)-(54) \begin{equation}
\tilde{T}^{00}=\frac{1}{(2\pi)^{3}}\kappa^{2}(\theta+A)a^{-4}
\Big(24\pi (1+c_{0}(1))+96\pi a^{-1}\tilde{c}_{1}(1)\Big)
\end{equation} and for (55)-(57)\begin{equation}\begin{array}{l}
 \tilde{T}^{00}=\frac{1}{(2\pi)^{3}}\kappa^{2}\theta^{-1}(\theta+A)a^{-4}
\Big(24\pi(1+c_{0}(2))+96\pi a^{-1}\tilde{c}_{1}(2)\cr +480\pi
\tilde{c}_{2}(2)a^{-2}\Big)\end{array}
\end{equation}
The model (85) leads (according to eq.(73)) to the Friedman
equation
\begin{equation}\begin{array}{l}(a^{-1}\frac{da}{d\tau}
)^{2}=\sigma(A+\theta)a^{-4}-\sigma\int_{\tau_{0}}^{\tau} dra^{-3}
+\frac{\Lambda}{3}(\tau_{0})\end{array}\end{equation} where
\begin{equation}\sigma=\frac{1}{(2\pi)^{3}}48
G\pi^{2}\kappa^{8}\theta^{3}
\end{equation}
 We can find an
explicit power-like solution of the integro-differential equation
(88) by a fine tuning of parameters showing that the exponential
behaviour (77) is not a necessity even if $\Lambda(\tau_{0})>0$.
Let us assume
\begin{equation} a(\tau)=\nu (\tau-q)^{\gamma}
\end{equation}
with the initial condition $a(\tau_{0})=\nu(\tau_{0}-q)^{\gamma}$.
Inserting (90) into eq.(88) we determine the parameters in eq.(90)
\begin{equation}
\gamma=1,
\end{equation}
\begin{equation}
\nu=\sigma^{\frac{1}{3}},
\end{equation}\begin{equation}
(\tau_{0}-q)^{2}=\frac{2\theta}{\nu}
\end{equation}
and \begin{equation}
\Lambda(\tau_{0})=\frac{3}{2}(\tau_{0}-q)^{-2}.
\end{equation}We can calculate the rhs of the Friedman equation
which is $\frac{1}{3}(8\pi G\tilde{E}+\Lambda)$. We obtain
\begin{equation}\Lambda=8\pi G
\tilde{E}=\frac{3}{2}(\tau-q)^{-2}\end{equation} Eq.(95) means
that $E-3\Pi_{E}=0$ in agreement with the discussion at eq.(78)
(there is no deceleration).

Eq.(90) applies  if $q<\tau_{0}$ because the integral in eq.(88)
is divergent at $r=q$.  The solution (90) defined on the interval
$[\tau_{0},\infty)$ does not achieve  $0$ reaching only its
minimum at $a(\tau_{0})=\nu(\tau_{0}-q)$. The solution (90) with
$\gamma=1$  is interesting because it gives $H^{-1}$ (where $H$ is
the present value of the Hubble constant) as the age of the
Universe in agreement with recent experimental data
(\cite{cheng},sec.11.4.1; see also \cite{nano} \cite{flavio} for
an explanation of a distinguished character of the linear
evolution).

 For general
$\tilde{T}^{00}$ resulting from eq.(40) we are unable to solve
eq.(73) exactly. We can obtain some estimates on the asymptotic
behaviour. We can have an exponential or power-like evolution
depending on the parameters entering eq.(73). For a discussion of
asymptotic solutions it is useful to differentiate the
integro-differential equation (73). We obtain
\begin{equation}
\frac{d}{d\tau}(a^{-1}\frac{da}{d\tau} )=-\frac{16\pi
G}{3}\tilde{T}^{00}.\end{equation} Note that differentiation of
the standard Friedmann equation (64) (with $\tilde{T}^{00}\simeq
a^{-4}$) also gives eq.(96) but this coincidence is accidental. It
results from the power -4 in $a^{-4}$. Nevertheless, if
$\tilde{T}^{00}\simeq a^{-r}$ with $r>0$ then there is no
essential difference between the standard Friedmann equation and
(96) up to an undetermined cosmological constant which is
anihilated when we differentiate the standard Friedmann equation.
Eq.(96) can be rewritten as the third order equation for $A$ (with
$A^{\prime}=\frac{dA}{d\tau}=a$)
\begin{equation}\begin{array}{l}
A^{\prime\prime\prime}-(A^{\prime\prime})^{2}(A^{\prime})^{-1}
=-\frac{16\pi G}{3}T^{00}(A,A^{\prime})A^{\prime}.
\end{array}\end{equation}
It can be seen that the $a^{-n}$  terms in eq.(84) do not modify
the large time behaviour because they decay faster in time. The
Ansatz $A\simeq \tau^{\gamma +1}$ in eqs.(82)-(84)and (97) for a
large time gives again $\gamma$ of eq.(91).

For the early Universe we apply the finite temperature solutions
of the diffusion equation of sec.5. We are interested in the small
time behaviour of the solutions of the finite temperature
diffusion on the time interval $[0,\tau_{0})$. Using eq.(97) we
can show that the polynomial modification (40) of the J\"uttner
initial condition changes the small time behaviour of the
solutions of the Friedmann equations. At $a\simeq 0$ the most
singular terms in $a$ dominate in eq.(84). Inserting the
energy-momentum tensor (84) in the Friedmann equations (97) we
obtain
\begin{equation}\begin{array}{l}
A^{\prime\prime\prime}-(A^{\prime\prime})^{2}(A^{\prime})^{-1}
\simeq -C(A^{\prime})^{-N-3}
\end{array}\end{equation}with a certain constant $C$.
Eq.(98) leads to the behaviour
\begin{displaymath}
a\simeq \tau^{\frac{2}{4+N}}
\end{displaymath} when $\tau\rightarrow 0$.
Hence, with an increasing $N$ the approach to the singularity
$a=0$ becomes  slower than in the exact solution (90) for $\tau>q$
and also slower than the  $\sqrt{\tau}$ behaviour of the solution
of the standard Friedmann equation (64). If the dependence of $T$
on $a$ is more singular than power-like then  we can encounter
solutions of the Friedmann equation (73) which do not reach $a=0$.

The solutions of secs.3 and 4 could also describe an
ultrarelativistic expanding ball,e.g., an exploding star or a
hadron (fireball) formed after a heavy ion collision
\cite{rafelski}\cite{svet}. In such a case we are interested in an
equation of state of such an object. Applying the formula (60) and
the relation between $\tilde{E}$ and $\tilde{\Pi}_{E}$ (68) we can
derive the relation between the pressure $\tilde{\Pi}_{E}$,
density $n$ and temperature $(\beta a)^{-1}$. For the
energy-momentum (82) we obtain
\begin{equation}\begin{array}{l}
\tilde{\Pi}_{E}=n(\beta
a)^{-1}\Big(1+c_{0}(1)+4\tilde{c}_{1}(1)(\beta
a)^{-1}\Big)\Big(1+c_{0}(1)+3\tilde{c}_{1}(1)(\beta
a)^{-1}\Big)^{-1}.\end{array}
\end{equation}
Eq.(99) modifies  the well-know equilibrium relation
$\tilde{\Pi}_{E}=n(\beta a)^{-1}$ true (according to eq.(99)) for
low temperatures to $\tilde{\Pi}_{E}=\frac{4}{3}n(\beta a) ^{-1}$
at high temperatures. The coefficient in the high temperature
limit depends on $N$ in the states (84).
\section{Non-relativistic diffusive fluid}
In sec.5 we have studied an ultrarelativistic diffusion (21) at
finite temperature for a hot matter. Then, in sec.6 the evolution
of $\Omega$ continued as an ultrarelativistic diffusion at zero
temperature. Finally, in this section we consider Universe
evolution in the stage when matter is in the form of macroscopic
bodies (we treat it as a dust of heavy particles). We still assume
that the Universe evolution is described by the diffusion equation
(17) in the limit of a large mass $m$. This is at the same time
the non-relativistic limit of eq.(17). The diffusion constant
$\kappa$ at this stage of evolution is different then the one for
the early Universe but we keep the same symbol for it. The
diffusive behaviour of a particle (dust) could be related to the
gravitational attraction of all other masses in the Universe
\cite{chandra2}. In this section we find solutions of the
non-relativistic diffusion equation, calculate the corresponding
energy momentum tensor $\tilde{T}$, derive the formula for $T_{D}$
and discuss the resulting Einstein equations.

 In the non-relativistic limit only the massive
terms on the rhs of eq.(17) remain. We redefine the diffusion
constant $\kappa^{2}\rightarrow \frac{\kappa^{2}}{2m}$ in eq.(17)
and write the equation in terms of the cosmic time
$\partial_{t}=a\partial_{\tau}$. Then, the non-relativistic limit
of eq.(17) in the cosmic time reads
\begin{equation}
\partial_{\tau}\Omega-2H p\frac{\partial \Omega}{\partial
p}=\frac{\kappa^{2}}{2a^{2}}\Big(\frac{\partial^{2} \Omega}{\partial
p^{2}} +\frac{2}{p}\frac{\partial \Omega}{\partial p}\Big),
\end{equation}where $H=a^{-1}\frac{da}{d\tau}$ (eq.(18)). We proceed as in sec.4
 where the solution began to evolve from the
J\"uttner equilibrium  distribution. We consider a solution of the
diffusion equation starting from the Maxwell distribution at
$\tau=\tau_{0}$\begin{equation}
 \Omega_{0}=n a^{3}(\tau_{0})(2\pi)^{3}(2\pi\kappa^{2})^{-\frac{3}{2}}\theta^{-\frac{3}{2}}
\exp\Big(-\frac{a^{4}(\tau_{0})p^{2}}{2\theta\kappa^{2}}\Big).\end{equation}
So, $ \frac{\theta\kappa^{2}}{a^{4}(\tau_{0})m}$ has a meaning of
the temperature. The state $\Omega_{0}$ is normalized to the
particle density $n$, i.e.,
\begin{displaymath}
\sqrt{h}(\tau_{0})\int\frac{ d{\bf
p}}{(2\pi)^{3}}\Omega_{0}=n.\end{displaymath} The solution of
eq.(100) with the initial condition (101) reads
\begin{equation} \Omega_{\tau}=na^{3}(\tau_{0})(2\pi \kappa^{2})^{-\frac{3}{2}}(\theta+A_{d})^{-\frac{3}{2}}
\exp\Big(-\frac{a^{4}(\tau)p^{2}}{2(\theta+A_{d})\kappa^{2}}\Big)\end{equation}
where
\begin{equation}
A_{d}(\tau)=\int_{\tau _{0}}^{\tau}ds a^{2}(s ).
\end{equation}
Note that in contradistinction to $A$ in eq.(28) now the integral
in $A_{d}$ is performed over the cosmic time. The energy density
in the state (102) is
\begin{equation}\begin{array}{l}
\tilde{T}^{00}=\sqrt{h}\int\frac{ d{\bf
p}}{(2\pi)^{3}}\sqrt{m^{2}+a^{2}p^{2}}\Omega_{\tau}\cr\simeq
\sqrt{h}\int \frac{ d{\bf
p}}{(2\pi)^{3}}(m+\frac{1}{2m}a^{2}p^{2})\Omega_{\tau}=nma^{3}(\tau_{0})a^{-3}(\tau)+C\kappa^{2}(\theta+A_{d})a^{-5}
\end{array}\end{equation}with certain positive constants  $C$ (independent of $\kappa$). In eq.(104)
 (as usual) we have cut the non-relativistic approximation of the energy at
the second order term.
 We can now
repeat the derivation of the modified Friedmann equation (73) with
the non-relativistic matter corresponding to $\tilde{\pi}_{E}=0$
in eq.(71) (the "dust") leading to\begin{equation}\begin{array}{l}
\frac{3}{8\pi G}(a^{-1}\frac{da}{d\tau}
)^{2}=\tilde{T}^{00}-\int_{\tau_{0}}^{\tau}dra^{-3}\partial_{r}(a^{3}\tilde{T}^{00})
+\frac{\Lambda}{8\pi G}(\tau_{0}).\end{array}\end{equation}
Inserting the result (104) for $\tilde{T}^{00} $ we obtain the
Friedmann equation
\begin{equation}\begin{array}{l}\frac{3}{8\pi
G}(a^{-1}\frac{da}{d\tau}
)^{2}=nma^{3}(\tau_{0})a^{-3}(\tau)+C\kappa^{2}(A_{d}+\theta)a^{-5}\cr-C\kappa^{2}\int_{\tau_{0}}^{\tau}
dra^{-3}
+2C\kappa^{2}\int_{\tau_{0}}^{\tau}H(\theta+A_{d})a^{-5}dr+\frac{\Lambda}{8\pi
G}(\tau_{0})\end{array}\end{equation} Depending on the parameters
in eq.(106) it  has the asymptotic solution for a large time
\begin{displaymath}
a(\tau)\simeq\tau
\end{displaymath}
besides the exponentially growing solutions (77). If there is no
diffusion ($\kappa=0$) then eq.(106) is reduced to the standard
Friedmann equation for a cosmological dust.

\section{Summary and outlook}In our approach
we replace the Einstein-Vlasov  system investigated by many
authors by a dissipative system.The usual form of the $a^{-4}$
dependence of the energy density on the scale factor for an
ultrarelativistic fluid can be considered as a result of the
J\"uttner (or Planck) equilibrium phase space distribution with a
time-dependent temperature. It also follows from the
energy-momentum conservation of an ideal fluid with the
 density-pressure relation for massless particles.
In this paper we considered phase space distributions resulting
from a relativistic diffusion. The diffusion approximation comes
out from a general assumption of the Markov property of
microscopic interactions. There are some general restrictions on
the form of the diffusion resulting from the relativistic
invariance and the detailed balance. They are independent of the
form of the interaction. The dissipative dynamics neglects some
degrees of freedom (in astrophysical context:some forms of
matter). For this reason the energy-momentum of the diffusing
fluid is not conserved. To some extent we can recover the missing
part of the total system from the requirement of the
energy-momentum conservation.

We have studied a simplified model of a scenario for the Universe
evolution which begins with a diffusion at finite temperature with
the Planck or  J\"uttner equilibrium distribution. At later time
(lepton era) a description in terms of an equilibrating diffusion
cannot apply. However, assuming that the dynamics remains
dissipative we continue with the Markov approximation describing
the evolution of the phase space distribution by relativistic
diffusion (with a different diffusion constant) without any
equilibrium heat bath. The corresponding energy-momentum tensor is
not conserved. In order to achieve the conservation law for the
total energy-momentum we introduce a "cosmological term". Under
the assumption of homogeneity of the (dark) matter distribution
the cosmological term is determined by the form of the
energy-momentum of the diffusing fluid. We have assumed that
particles are massless. This assumption can be justified at an
early stage of Universe evolution either as an ultrarelativistic
approximation or referring to the grand unified theory
 in which masses arise at lower temperatures through the symmetry breaking.
 At the
later stage the massless fluid must be considered as a
mathematical simplification (21) of eq.(17) which allows to solve
the diffusion equation in order to derive the form of the
energy-momentum tensor as a function of the scale factor.
Inserting the solution of the diffusion equation into the Einstein
equations we can derive a dissipative version of the Friedmann
equation. We have studied some consequences of this modification
of the Friedmann equation. We show that the cosmological term
starting from the cosmological constant (as a free integration
parameter) is decreasing with time. If it tends to zero then the
Friedmann equation has a solution of the form $a\simeq \tau$ for a
large time. If the cosmological constant is large enough then the
Friedmann equation has an exponentially growing solution. We would
need to apply numerical methods to study the scale factor
evolution in order to see whether it can support a fast expansion
at an intermediate time simulating the inflation. At the later
stage of Universe evolution when matter acquires a macroscopic
form we describe the evolution by a non-relativistic diffusion of
a dust of heavy particles. We show that such an evolution leads to
the same conclusions concerning the long time behaviour as the
relativistic diffusion at zero temperature. We have proposed a
simplified model of the Universe evolution which has a built in
equilibrium stage and allows explicit solutions. The approach
could be considered as an approximation of the collision term in
the Boltzmann equation (treated, e.g., in \cite{dodelson}) by a
diffusion applied to the case when some scattering processes
causing the energy dissipation remain unknown. It gives a
phenomenological description of quantum processes in an early
Universe. We were able to study some consequences of our
assumptions. In particular, it could be seen that the dissipation
determines the large time behaviour of $a(\tau)$. The early times
of the Universe dynamics are accessible to astronomical
observations through their impact on the structure formation. We
intend to continue the study by an admission of a non-homogeneity
of the metric in order to investigate the result of the diffusive
dynamics of the hot Universe on  the propagation of disturbances.
 For a small time the
modified J\"uttner solutions of the diffusion equations at finite
temperature lead to a slower approach to the $a=0$ singularity of
the scale factor. Such a behaviour could possibly prevent the
appearance of the singularity at $\tau=0$ in a more general
dissipative dynamics. The admittance of polynomial modifications of
the J\"uttner distribution (as solutions of the diffusion equation)
leads to a modification of the equation of state defined as a
relation between the density, temperature and pressure.

{\bf The acknowledgements}

The author thanks an anonymous referee for inspiring questions.
Interesting discussions with Andrzej Borowiec and Marek Szydlowski
are gratefully acknowledged. The research is supported by Polish
NCN grant No.2013/09/B/ST2/03455.

\end{document}